\documentclass[useAMS,usenatbib,color]{mn2e}
\usepackage{natbib}
\usepackage{amssymb,amsmath,amsfonts}
\usepackage{epsfig}

\title{ The minimum stellar metallicity observable in the Galaxy}

\author[A. Frebel, J.L. Johnson and V. Bromm]
{Anna Frebel\thanks{E-mail: anna@astro.as.utexas.edu}, Jarrett L. Johnson and Volker Bromm \\
McDonald Observatory and Department of Astronomy, University of Texas, Austin, TX 78712, USA \\}

\pubyear{2008}

\begin{document}
\maketitle
\topmargin-1cm

%\label{firstpage}
\begin{abstract}
The first stars fundamentally transformed the early Universe through
their production of energetic radiation and the first heavy chemical
elements. The impact on cosmic evolution sensitively depends on their
initial mass function (IMF), which can be empirically constrained
through detailed studies of ancient, metal-poor halo stars in our
Galaxy. We compare the lowest magnesium and iron abundances measured
in Galactic halo stars with theoretical predictions for the minimum
stellar enrichment provided by Population\,III stars under the
assumption of a top-heavy IMF. To demonstrate that abundances measured
in metal-poor stars reflect the chemical conditions at their
formation, and that they can thus be used to derive constraints on the
primordial IMF, we carry out a detailed kinematic analysis of a large
sample of metal-poor stars drawn from the SDSS survey. We assess
whether interstellar accretion has altered their surface
abundances. We find that accretion is generally negligible, even at
the extremely low levels where the primordial IMF can be tested. We
conclude that the majority of the first stars were very massive, but
had likely masses below $\sim 140\,M_{\odot}$.

\end{abstract}

\begin{keywords}
cosmology: early Universe ---
stars: abundances ---
stars: kinematics ---
stars: Population II ---
Galaxy: halo ---
techniques: spectroscopic
\end{keywords}

\section{Introduction}
The first stars, the so-called Population\,III (Pop\,III), were the
key drivers of early cosmic evolution. Their copious production of
hydrogen-ionizing radiation initiated the reionization of the
Universe, and the first supernova (SN) explosions seeded the pristine
intergalactic medium (IGM) with the first heavy elements
\citep{ciardi, barkana_loeb}. The character of this stellar feedback
sensitively depends on the initial mass function (IMF) of the first
stars. The current theoretical model of their formation, based on
numerical simulations, suggests that the Pop\,III IMF was top-heavy
\citep{brommARAA}. In the context of modern cold dark matter (CDM)
cosmology, there are two physically distinct sites of early star
formation. The very first stars are predicted to have formed in
so-called minihaloes at redshift $z\sim 30 - 20$. The subsequent SNe
dispersed the first heavy elements into the surrounding gas, thus
setting the initial conditions for the formation of the
second-generation of already slightly metal-enriched (Pop\,II) stars.
If the prediction of a top-heavy IMF is correct, of order one Pop\,III
star would form per minihalo, whereas a small cluster of predominantly
low-mass Pop\,III stars would arise in the alternative case of a
normal, Salpeter-like IMF. The second site for the formation of stars
at high redshift are the atomic cooling haloes, with of order a 100
times the mass of the minihaloes. Their dark matter potential wells
are sufficiently deep to induce the collapse of the material that was
affected by the SN feedback from the Pop\,III stars in
minihaloes. These systems are therefore the sites for the formation of
the second generation (Pop\,II) stars. Due to their predominantly low
masses they may still be found today as the most metal-poor stars.

Each galaxy thus exhibits a certain minimum metallicity, reflecting
the pre-enrichment from Pop\,III stars. We here explore the
fundamental question of the minimum Fe and Mg abundances observable in
the Milky Way to derive constraints on early galaxy formation and on
the Pop\,III IMF. We pursue a ``near-field cosmology'' approach
\citep{blandhawthorn_freeman}, established over the past decade by
large objective-prism surveys \citep{ARAA} of metal-poor stars as
vital tracers of Galactic chemical evolution and the early Universe.
These stars carry the fossil record of the physical conditions in the
first star forming systems (``stellar archaeology''). To successfully
retrieve any such signatures, it is crucial that the atmospheric
composition of the observed stars has not been altered either
intrinsically by the products of nuclear burning in the stellar
interior, or externally by accretion of interstellar material during
their long lifetimes. Mass transfer across a binary system may also
change the abundances of certain elements (e.g. C), but not the ones
that are of interest in this study (Fe, Mg).  
The first effect can be accounted for by selecting relatively
unevolved main-sequence and giant stars.  Regarding the second issue,
only a few approximate calculations are available
\citep{talbot,yoshii81,Iben1983}, based on the idealized assumption
that all stars have the same velocity.  We therefore
revisit the issue of accretion with a full stellar kinematic analysis
of a large sample of metal-poor stars, so that we can assign
individual velocities to them. A more realistic
modeling of accretion is crucial because stellar archaeology
pre-supposes a negligible contribution to the observed abundances from
such pollution, so that it is possible to derive constraints on the
early Universe and the Pop\,III IMF. Testing the prediction of a
top-heavy IMF is one of the main goals of the upcoming {\it James Webb
Space Telescope (JWST)}, but it is important to also utilize
complementary probes that are already accessible now, such as the most
metal-poor stars.

\section{Minimum Stellar Metallicity}
A fundamental characteristic of the Milky Way is the minimum
observable metal-enrichment in its stars. The existence and level of
such a ``metallicity floor'' is governed by the Pop\,III IMF.  If the
first stars were formed with a normal, Salpeter-like IMF, contrary to
the current consensus view, there would be no minimum stellar
metallicity, and truly metal-free, low-mass stars would exist. In this
case, significant interstellar accretion could masquerade such
putative primordial abundances in those stars. Without detailed
knowledge of their accretion history, this would prevent us from
identifying them as such low-mass Pop\,III stars. Hence, any
information about the IMF would be irretrievably lost.

\begin{figure}
\begin{center}
\includegraphics[clip=true,bbllx=34,bblly=205,bburx=252,bbury=550, scale=]{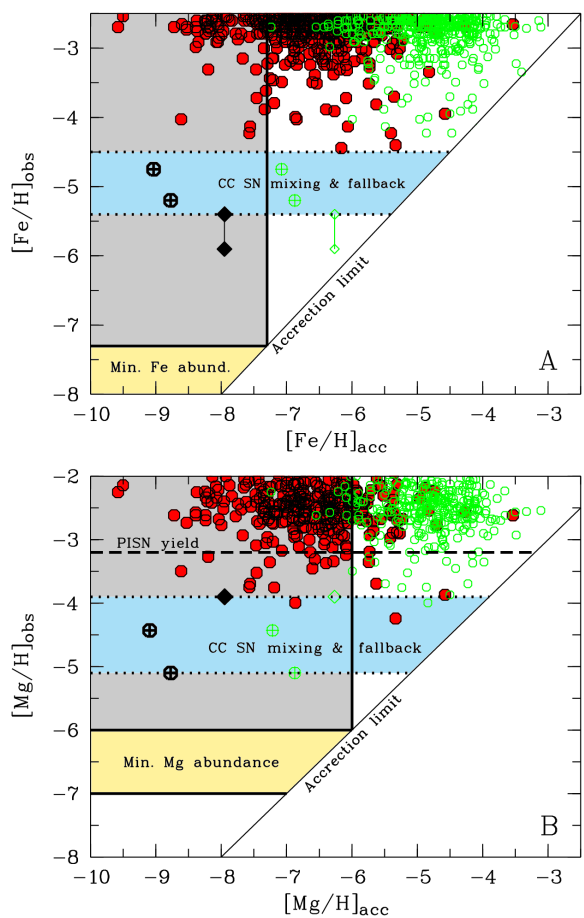}
  \caption{\label{fig1} Observed vs. ``accreted'' Fe ({\it panel A})
and Mg ({\it panel B}) abundances for the sample of metal-poor stars
({\it red circles}). The case where all stars pass through a dense
cloud once is also presented ({\it green open circles}). The three
most Fe-poor stars are indicated ({\it crossed circles, diamond}). For
HE~1327$-$2326, both the 1D~non-LTE and 3D~LTE Fe values are
shown. All stars have accreted fewer metals than what is observed,
thus demonstrating the validity of the basic assumption underlying
stellar archaeology. The minimum Fe and Mg abundance ranges,
calculated under the assumption of a top-heavy Pop\,III IMF, are given
({\it yellow region}). The approximate Mg abundance arising from a
PISN event in an atomic cooling halo is indicated ({\it dashed line,
panel B}), as well as the Fe and Mg levels of enrichment from a
$25\,M_{\odot}$ mixing and fallback SN ({\it dotted lines, blue
regions}).  We highlight those stars ({\it gray regions}) that can be
used to place constraints on the Pop\,III IMF, where accretion does
not affect whether they lie above or below the theoretical ``bedrock
abundances'', predicted for a top-heavy IMF. Since all observed Mg
abundances within the IMF-sensitive ({\it gray}) region to date fall
above this bedrock range, we conclude that a top-heavy IMF is favored
for the first stars.}
\end{center}
\end{figure}

For a top-heavy IMF, the situation is very different.  Recent
numerical simulations of the assembly process of atomic cooling
haloes, which are often thought to constitute the first galaxies, have
shown that Pop\,III star formation only occurs in a few of the
progenitor minihaloes that eventually merge into the atomic cooling
halo \citep{johnson08}. For simplicity, we here assume that only one
minihalo hosted a Pop\,III star that ended its life in a SN
explosion. This accommodates the possibility that a fraction of the
minihalo Pop\,III stars formed massive black holes by direct collapse,
without any concomitant metal enrichment. Under this assumption, we
can now derive an estimate for the ``bedrock enrichment'' from
Pop\,III stars with a top-heavy IMF, which would in turn set the
minimum stellar metallicity observable in the Galaxy's oldest Pop\,II
stars. Current simulations predict the typical Pop\,III mass to only
within a factor of 10. Consequently, within the general top-heavy
paradigm, a number of qualitatively very different SN pathways for the
first stars are still possible \citep{heger2002,iwamoto_science}, and
it is important to consider these. If the progenitor Pop\,III star had
a mass in the range $140-260\,M{_\odot}$, an energetic
pair-instability SN (PISN) would occur, which is characterized by
extremely large metal yields \citep{heger2002}. The Mg yield is almost
constant over the entire PISN mass range; assuming that the Mg yield
from a single PISN is well-mixed in an atomic cooling halo containing
a total gas mass of $10^7\,M{_\odot}$ leads to the narrowly confined
prediction of $\mbox{[Mg/H]}_{\rm min} \simeq -3.2$.\footnote{
\mbox{[A/B]}$=\log_{10}(N_{\rm A}/N_{\rm B}) - \log_{10}(N_{\rm
A}/N_{\rm B})_\odot$, for elements A, B.}  Since this overlaps with
the range of observed stellar Mg abundances, a fraction of Pop\,II
stars could carry the signature of PISN nucleosynthesis.  However,
this fraction is likely very small since no clear PISN ``odd-even''
effect has thus far been found among metal-poor stars. No useful PISN
prediction can be made in the case of Fe, since the corresponding
yields range from zero to very high values, depending on the precise
progenitor mass \citep{heger2002}.  Alternatively, if the Pop\,III
progenitor had a less extreme mass, but still in the black hole
forming range of $M_{\ast}\ga 25 M{_\odot}$, a peculiar class of
``faint'', core-collapse (CC) SNe becomes possible.  The class of such
low explosion-energy SNe, experiencing mixing and fallback onto a
nascent black hole, was introduced to produce very low Fe yields
\citep{UmedaNomotoNature,iwamoto_science} in order to explain the two
hyper Fe-poor stars with $\mbox{[Fe/H]}<-5.0$.  Higher explosion
energies are required to explain the abundance pattern of metal-poor
stars with $\mbox{[Fe/H]}\gtrsim-4.5$.  We use these observationally
calibrated nucleosynthesis calculations to constrain the likely range
in Fe and Mg abundances that would result if the first stars died as
such faint CC SNe. The different pre-enrichement levels are indicated
in Fig.~\ref{fig1}, for comparison with the observational data.

Based on these SN yield considerations, we have derived typical values
for the Pop\,III pre-enrichment. We now wish to place extreme lower
limits on the observable stellar Mg and Fe abundances in the Galaxy
that result from assuming a top-heavy Pop\,III IMF. It is often
argued that the transition from a top-heavy to a normal, Salpeter-like
IMF for the subsequent generations of Pop\,II/I stars (including those
considered here) is governed by a ``critical metallicity''
\citep{brommARAA}. Its value is still rather uncertain, depending on
whether fine-structure line cooling is dominant
\citep{brommnature,dtrans}, or dust cooling \citep{schneider06}. To
arrive at a robust estimate that does not depend on the detailed
nature of the critical metallicity, we consider a range that extends
from typical fine-structure to dust predictions. Within the
fine-structure model \citep{brommnature}, carbon is the most important
coolant, leading to a critical abundance of $\mbox{[C/H]}_{\rm
min}=-3.5$. Combining this with the empirically determined maximum
carbon-to-magnesium and carbon-to-iron ratios found in metal-poor
stars, i.e. in HE~0107$-$5240 ($\mbox{[C/Mg]}_{\rm max}=2.5$;
\citealt{collet06}) and HE~1327$-$2326 ($\mbox{[C/Fe]}_{\rm max}=3.8$;
\citealt{he1327_uves}), we estimate the minimum Mg and Fe abundances
to be $\mbox{[Mg/H]}_{\rm min}=-6.0$ and $\mbox{[Fe/H]}_{\rm
min}=-7.3$. Dust cooling models typically result in lower critical
abundances, $\mbox{[C/H]}_{\rm min}=-4.5$, and the minimum observable
Fe and Mg abundances are reduced accordingly.  Our predictions for the
minimum Fe and Mg values are shown in Fig.~\ref{fig1} ({\it yellow
regions}).

The level of our predicted metallicity floor is particularly
interesting for the goals of current and future surveys with regard to
identifying the most metal-poor stars. Some of the recently discovered
metal-poor stars have extremely low Mg and Fe abundances that begin to
approach our theoretical predictions for the metallicity floor.  These
objects suggest that additional examples of such stars can be found
with current observational techniques, potentially even with
abundances below the current record holders. Based on spectrum
synthesis calculations, we estimate that suitably cool giants should
have at least one detectable Mg and Fe line at abundances as low as
$\mbox{[Mg/H]}\sim-6.5$ and $\mbox{[Fe/H]}\sim-8.0$,
respectively. Technological limitations should therefore not prevent
us from reaching abundances that are within our predicted minimum
metallicity ranges.

\section{Data on Metal-Poor Stars}
In order to explore whether the most metal-poor stars used here to
constrain the minimum stellar abundances, or other metal-poor halo
stars in general, are possibly affected by accretion, we reconstruct
their individual accretion histories by carrying out a detailed
kinematical analysis. Our sample stars are selected from the Sloan
Digital Sky Survey (SDSS), which provides the necessary input data
(radial velocities, distances, proper motions, abundances) for such an
analysis.\footnote{See http://www.sdss.org/dr6/.} Studies based on the
kinematics from SDSS have already led to groundbreaking results
(e.g. \citealt{carollo}). A full description of the data products used
here can be found elsewhere \citep{usno-b,lee_sspp1,dr6}. We note
though that we employed [Fe/H] abundances derived from the Ca\,K line
and selected 565 stars with $\mbox{[Fe/H]}<-2.5$. The spectra of all
stars with $\mbox{[Fe/H]}<-3.4$ were inspected because the majority of
them turned out to be spectral artifacts or misclassified objects, and
not real metal-poor stars. This leaves 474 stars in the sample. To
obtain [Mg/H], we set the available [$\alpha$/Fe] equal to [Mg/Fe].
Based on temperature estimates from the H\,$\delta$ line, we find that
the sample contains 472 turnoff (dwarf) stars with known proper
motions; the remaining two appear to have unrealistically low
temperatures so we exclude them from the sample.  Reliable proper
motions are not available for most of the well-studied metal-poor
giants since their distances are very large, and hence
uncertain. Future missions such as GAIA will enable us to extend this
work by providing accurate proper motions, especially for all the
metal-poor giants. Since our diagnostic is based on readily obtainable
medium-resolution spectra, it will be straightforward to apply it to
the extensive data sets from future large-scale surveys.

\section{Role of Accretion}
We assume that a given star moves in a rigid, three-component Milky
Way potential, adopted from \citet{johnston98}, for 10\,Gyr. Using a
standard orbit integrator (D. Lin, priv. comm. 2008; see
\citealt{fulbright} for further details), we determine the orbital
parameters, such as $U, V, W$ velocities and eccentricity, for all
sample stars. For simplicity, accretion is assumed to take place only
during disk crossings.  The density structure in the disk interstellar
medium (ISM) is highly inhomogeneous, such that every star will
encounter regions of different density at each disk crossing. Since
the accretion rate depends only linearly on density (see Sec.~4), as
opposed to the inverse-cubed scaling with velocity, we here for
simplicity work with an average ISM density. Using the empirically
determined volume filling fraction as a function of density
\citep{talbot}, we find for the average disk density, $n\simeq
5$\,cm$^{-3}$ and assume a disk height of $\sim 100$\,pc.

We estimate the amount of accreted gas onto a
low-mass star that passes through interstellar gas assuming
Bondi-Hoyle accretion \citep{bondi}:
\begin{equation}
\dot{M}\simeq 2\pi
(GM)^{2}\rho/(v_{\rm rel}^{2}+c_{\rm s}^{2})^{3/2},
\end{equation}
where $M$ is the mass of the star, $\rho\simeq m_{\rm H} n$ the gas
density, $v_{\rm rel}$ the stellar velocity relative to the gas, and
$c_{\rm s}\simeq 5$\,km\,s$^{-1}$ the sound speed in the general Milky
Way ISM. We calculate the relative velocity of a star during each disk
crossing according to $v_{\rm rel} = \sqrt{(U, V-V_0, W)^2}$, where
$V_0\simeq 200$\,km\,s$^{-1}$ is the average rotation speed of the
disk. For the ISM abundances, we assume a solar distribution with the
overall metallicity evolving according to:
\begin{equation}
Z_{\rm ISM}(t) = \left(1 + t/t_{\rm H}\right)^{-4} \left[10^{-3} + 0.67 * \left\{(1 + t/t_{\rm H})^5 - 1\right\} \right] Z_{\odot},
\end{equation}
where t$_{\rm H}=13.7$\,Gyr is the Hubble time, such that $Z_{\rm ISM}
\sim Z_{\odot}$ at the time of the formation of the Sun $\sim$ 5\,Gyr
ago. This relation follows from detailed homogeneous chemical
enrichment models \citep{pagel}.  As halo stars will pass through the
Milky way disk between 50 and 80 times, the average density and
metallicity of the accreted gas is likely to be similar for all
stars. However, the stellar velocities at disk crossing can vary
widely, thus dominating the accretion rate because of the strong
dependence on relative velocity in equ.~(1).  Stars with the highest
velocity relative to the disk will experience the least pollution by
the Milky Way ISM, and are therefore most likely to display a surface
metallicity endowed at the earliest epochs of star formation. 

We specifically choose to investigate the accretion history of Mg and
Fe, which are easily measurable abundances in metal-poor stars. Also,
these abundances are not affected by potential mass transfer across a
binary system. We calculate the total amount of accreted Mg and Fe by
summing up the contributions from every disk crossing. To finally
arrive at surface abundances, we assume that $\sim 10^{-3}$ of the
stellar mass is contained in the convective outer layer for
dwarfs. For giants this fraction would rise to $\sim 0.1$
\citep{yoshii81}.

In Fig.~\ref{fig1}, we compare the resulting accreted Mg and Fe
abundances for every star (all dwarfs) with its observed values. All stars have
lower ``accreted abundances'' than their observed abundances. This
generally confirms that accretion does not dominate the abundance
pattern of old metal-poor stars, thus rendering the overall concept of
``stellar archaeology'' viable. It also suggests that if the
first stars were characterized by a Salpeter-like IMF, and if they traveled at
sufficiently high velocity, it should be possible to find
surviving stars with arbitrarily low abundances. At these low
metallicities, even traces of accreted material may have an impact on
the surface abundances, making detailed knowledge of any potential
accretion indispensable.

For the three most Fe-poor stars currently known, the subgiant
HE~1327$-$2326 ($\mbox{[Fe/H]}=-5.9$; \citealt{HE1327_Nature,
he1327_uves}), together with the giants HE~0107$-$5240
($\mbox{[Fe/H]}=-5.2$; \citealt{HE0107_Nature}) and HE~0558$-$4840
($\mbox{[Fe/H]}=-4.8$; \citealt{he0557}) we find the same
result. Their accretion levels are somewhat uncertain (by $\sim
1$\,dex) due to poorly determined distances and proper motions, but
still indicate an approximate level of accretion that is well below
their observed abundances. Regarding distance and proper motion
uncertainties for the other stars, we verified that the derived
amounts of accreted material are not significantly affected when the
input parameters are changed by up to 30\%.

We also consider the special case where each star passes
once through an extremely dense cloud of 100\,pc diameter and $n\sim
10^3$\,cm$^{-3}$, comparable to the inner region of a giant molecular
cloud (GMC).  We further assume that this occurs when the star has its
smallest $v_{\rm rel}$, to maximize the potential for accretion. In
this case, the observed abundance pattern may be dominated by the
signature of the ISM. It is not known whether every star encounters a
GMC, but even in the case it does, the accretion process will strongly
depend on the space velocity of the star. In Fig.~\ref{fig1}, we show
the total accreted Mg and Fe for all stars in this extreme case ({\it green
circles}). We find that the stars still have slightly
lower ``accreted abundances'' than the observed values. If a star were to
have an accreted abundance larger than the currently observed one, it
might indicate that such a star never entered a dense GMC during its
lifetime. Our GMC ``maximum accretion'' scenario thus provides a
robust upper limit to the total accreted abundance for each star. We
note that in our accretion estimates the potential role of stellar
winds has been neglected. The presence of a wind would likely balance any
accretion or prevent it altogether \citep{talbot}. Since wind strength
scales with stellar mass and metallicity, the low-mass, metal-poor
stars considered here should have little or no wind. Hence, the
accreted [Fe/H] and [Mg/H] abundances are likely an upper limit. We
thus conclude that stellar archaeology is not hampered by interstellar
accretion, even for our maximum accretion scenario.  Furthermore, we
demonstrate that kinematic information is vital for the identification
of the lowest-metallicity stars.

Finally, we compare our results with the ``pollution limit'' derived
previously by \citet{Iben1983}. This limit was calculated in an
attempt to explain the paucity of low-metallicity stars (G-dwarf
problem), within the framework of a normal, Salpeter-like IMF for the
first stars.  The estimated Fe pollution of $\mbox{[Fe/H]}_{\rm
acc}=-5.7$ (one value for all stars) would in this interpretation
naturally prevent the discovery of any stars with lower
metallicities. However, we show that the ``accretion limit'' is a
strong function of stellar kinematics, and that there is therefore no
such universal limit. Our result thus suggests that the traditional,
pollution-based, explanation for the absence of surviving low-mass
Pop\,III stars needs to be revisited. Furthermore, with the latest
Fe abundance measurement for HE~1327$-$2326 of $\mbox{[Fe/H]}=-5.9$
\citep{he1327_uves}, the Iben pollution limit has already been
reached. This star, however, is currently thought to be a second
generation object displaying the nucleosynthetic yields of a
metal-free CC SN with a mass $\sim25\,M_{\odot}$
\citep{iwamoto_science}, and not a masqueraded, low-mass Pop\,III
star. In addition, HE~1327$-$2326 does not exhibit scaled-down solar
abundances, contrary to the expectation that a star with an
accretion-dominated signature should show a solar abundance pattern.

\section{Implications}
As an interesting consequence from the preceeding analysis, we can
derive some observational constraints on the underlying Pop\,III IMF
that determined the level of pre-enrichment. From Fig.~\ref{fig1}, we
infer that select stars with high velocities and correspondingly low
accreted abundances ({\it gray region}) are useful probes of the
Pop\,III IMF. Accretion alone would not have been able to push them
above the minimum levels predicted for a top-heavy IMF.  If these
stars had formed from extremely low-metallicity gas, their observed
present-day surface abundances should still have reflected this.  The
fact that these low-accretion stars all have abundances above the
minimum floor predicted for a top-heavy Pop\,III IMF supports the
notion that the first stars were very massive (see also
\citealt{tumlinson2006,salvadori}). We here would like to repeat that such tests
can only be carried out if proper attention is given to the individual
accretion history of each star.

To make this test fully convincing, we need to address possible
observational biases. In particular, the apparent lack of
IMF-sensitive stars below the top-heavy prediction could simply
reflect their small numbers. However, current survey sizes reach
levels of completeness that render such an interpretation increasingly
unlikely. To gauge the putative number of low-mass Pop\,III stars in
the Galaxy, assuming a Salpeter-like primordial IMF, we begin with the
approximate number of minihaloes that formed before the redshift of
reionization, and that eventually merged to become part of the Milky
Way. Using standard extended Press-Schechter (EPS) theory
\citep{lacey}, we estimate that $\sim 10^{4}$ Milky Way progenitor
minihaloes hosted Pop\,III star formation.  The $\sim 100 M_{\odot}$
in cold, dense gas availale to form stars, as found in numerical
simulations, would then result in $\sim 100$ low-mass Pop\,III
stars. The total number of such hypothetical Pop\,III fossils in the
Galaxy would be $\sim 10^{6}$. Given that the Galactic halo today
contains $\sim 10^{9}$ stars, one low-mass Pop\,III star should be
found per $\sim 10^{3}$ stars surveyed \citep{oey2003}. In the
Hamburg/ESO survey, each of the two stars with $\mbox{[Fe/H]}<-5$ was
found in a sample of $\sim 2000$ selected metal-poor stars, which in
turn comprise $\sim10\%$ of a subset of halo stars with no metallicity
selection. To first order, it thus seems unlikely that a selection
effect would significantly affect our results. This argument is
further strengthend by SDSS, which has spectroscopically measured
metallicities for several hundred thousand stars.

Our results (see Fig.~\ref{fig1}, {\it panel B}) finally also suggest
that only a small fraction of the first stars died as PISNe because a
number of stars have observed [Mg/H] ratios below the abundance floor
predicted for PISN enrichment.  We thus conclude that the majority of
the first stars were very massive, but had masses below
$\sim140\,M_{\odot}$.  Current data and simulations are not yet
precise enough to determine the PISN fraction with any certainty,
but our diagnostic can in principle be extended to constrain this
important quantity (see also \citealt{karlsson08}). 

We have thus shown that stellar archaeology can provide crucial
observational constraints on the primordial IMF, given that the
metal-poor stars of interest have sufficiently high space velocities
to avoid significant accretion. Together with our prediction that
stars with abundances below the currently known lowest values can be found
in ongoing and future discovery efforts, stellar archaeology becomes
directly relevant for the science goals of the next generation of
30\,m-class optical telescopes such as the Giant Magellan Telescope (GMT)
and the Thirty Meter Telescope (TMT).
Future surveys will continue to provide us with
local constraints on star formation at the edge of the observable
Universe. Selecting suitable candidates for this task will inceasingly
rely on our ability to combine chemical abundance analyses with
kinematic information.

\section*{Acknowledgments} We thank A. Weiss, S. Cassisi and
H. Bluhm for helpful discussions. A.~F. acknowledges support through
the W.~J.~McDonald Fellowship of the McDonald Observatory. V.~B. is
supported by NSF grant AST-0708795. Funding for the SDSS
and SDSS-II has been provided by the Alfred P. Sloan
Foundation, the Participating Institutions, NSF,
the U.S. DoE, NASA,
the Japanese Monbukagakusho, and the Max
Planck Society, and the Higher Education Funding Council for England.
Further details can be found at http://www.sdss.org/collaboration/credits.html.

%============================================================================
%\bibliography{../BMPS/paperI.bib}
%\bibliographystyle{apj}

\end{document}